\documentclass[twocolumn]{IEEEtran}

\usepackage{amsmath,bm}
\usepackage{amsthm}
\usepackage{amssymb}
\usepackage{graphicx}
\usepackage{epsfig}
\usepackage{psfrag}
\usepackage{subfigure}
\usepackage{url}
\usepackage{stfloats}
\usepackage{amsmath}
\usepackage{algorithm}
\usepackage{algorithmic}
\usepackage{diagbox}
\usepackage{dsfont}
\usepackage{bbding}

\usepackage{amsmath}
\usepackage{graphics}
\usepackage{graphicx}
\usepackage{amssymb}
\usepackage{algorithm}
\usepackage{algorithmic}
\usepackage{mathrsfs}
\usepackage{booktabs}
\usepackage{bm}
\usepackage{bbm}
\usepackage{float}
\usepackage{url}
\usepackage{amsfonts}
\usepackage[square,sort&compress,numbers]{natbib}
\usepackage{multirow}
\usepackage{textcomp}
\usepackage{supertabular}
\usepackage{array}
\usepackage{color,soul}

\begin{document}

\title{Feedback Pyramid Attention Networks for Single Image Super-Resolution}

\author{ {Huapeng Wu, Jie Gui, \textit{Senior Member, IEEE}, Jun Zhang, \textit{Member, IEEE}, James T. Kwok, \textit{Fellow, IEEE}, and Zhihui Wei, \textit{Member, IEEE}}

\IEEEcompsocitemizethanks{
\IEEEcompsocthanksitem This work was supported in part by the National Natural
Science Foundation of China under Grant 11431015, Grant 61671243, Grant 61971223 and Grant 61572463. Corresponding author: Zhihui Wei (e-mail: gswei@njust.edu.cn).
\IEEEcompsocthanksitem Huapeng Wu is with the School of Computer Science and Engineering, Nanjing University of Science and Technology, 210094, China.
\IEEEcompsocthanksitem Jie Gui is with the School of Cyber Science and Engineering, Southeast University, 211100, China.
\IEEEcompsocthanksitem Jun Zhang is with the School of Science, Nanjing University of Science and Technology, 210094, China.
\IEEEcompsocthanksitem James T. Kwok is with the Department of Computer Science and Engineering, Hong Kong University of Science and Technology, Hong Kong.
\IEEEcompsocthanksitem Zhihui Wei (corresponding author: gswei@njust.edu.cn) is with the School of Computer Science and Engineering, Nanjing University of Science and Technology, 210094, China, and with the School of Science, Nanjing University of Science and Technology, 210094, China.

} 
} 

\markboth{}%
{Shell \MakeLowercase{\textit{et al.}}: Bare Demo of IEEEtran.cls for Journals}

\maketitle

\newcolumntype{L}[1]{>{\raggedright\arraybackslash}p{#1}}
\newcolumntype{C}[1]{>{\centering\arraybackslash}p{#1}}
\newcolumntype{R}[1]{>{\raggedleft\arraybackslash}p{#1}}

\begin{abstract}
Recently, convolutional neural network (CNN) based image super-resolution (SR) methods have achieved significant performance improvement. However, most CNN-based methods mainly focus on feed-forward architecture design and neglect to explore the feedback mechanism, which usually exists in the human visual system. In this paper, we propose feedback pyramid attention networks (FPAN) to fully exploit the mutual dependencies of features. Specifically, a novel feedback connection structure is developed to enhance low-level feature expression with high-level information. In our method, the output of each layer in the first stage is also used as the input of the corresponding layer in the next state to re-update the previous low-level filters. Moreover, we introduce a pyramid non-local structure to model global contextual information in different scales and improve the discriminative representation of the network. Extensive experimental results on various datasets demonstrate the superiority of our FPAN in comparison with the state-of-the-art SR methods.

\end{abstract}

\begin{IEEEkeywords}
Super-resolution, feedback mechanism, pyramid non-local structure.
\end{IEEEkeywords}

\section{Introduction}

Single image super-resolution (SISR) \cite{freeman2000learning} is a class of techniques that infers a high resolution (HR) image from its corresponding low resolution (LR) image. SR is an ill-posed inverse problem since each LR patch has to be mapped onto multiple HR patches. To solve this problem, researchers have proposed a multitude of learning-based methods to learn the mapping function from LR patches to their HR counterparts \cite{timofte2014a+, wu2018high, dong2015image, kim2016accurate, kim2016deeply, tai2017image, lai2017deep, 9208645, ledig2017photo, lim2017enhanced, 9395490, 9427111}.

Due to the strong learning ability, deep neural network \cite{lecun2015deep} based methods have been proposed to learn the nonlinear mapping between LR and HR image pairs. Dong \emph{et al}. \cite{dong2015image} firstly presented a super-resolution convolutional neural network (SRCNN) to learn an end-to-end nonlinear mappings from LR to its HR counterpart. Kim \emph{et al}. proposed a very deep super-resolution (VDSR) network \cite{kim2016accurate} by using residual learning with up to 20 convolutional layers, achieving substantial improvement over the SRCNN method. In order to reduce the parameters of the model, Kim \emph{et al}. further proposed Deeply-Recursive Convolutional Network (DRCN) \cite{kim2016deeply} via utilizing recursive-supervision and achieved similar results to VDSR. Tai \emph{et al}. introduced Deep Recursive Residual Network (DRRN) \cite{tai2017image}, which utilized local and global residual learning to train a deeper model and showed favorable performances against DRCN with fewer parameters. Based on Laplacian pyramid framework, Lai \emph{et al}. \cite{lai2017deep} proposed the Laplacian Pyramid Super-Resolution Network (LapSRN) to reconstruct HR images progressively. Ledig \emph{et al}. \cite{ledig2017photo} applied residual net (ResNet) architecture to construct a deeper network (SRResNet) for image SR. Enhanced Deep Super-Resolution (EDSR) \cite{lim2017enhanced} proposed by Lee \emph{et al}. employed a simplified ResNet \cite{he2016deep} architecture by removing the normalization layers in the SRResNet and won the championship of the NTIRE2017 super-resolution challenge \cite{timofte2017ntire}. To further improve the performance of the networks, Residual Dense Network (RDN) \cite{zhang2018residual} and Channel-wise and Spatial Feature Modulation Network (CSFM) \cite{hu2019channel} utilize the residual and the dense skip connections \cite{huang2017densely} in their networks to fuse multi-level features for the SR reconstruction.

\begin{figure*}
  \centering
  \includegraphics[width=\linewidth]{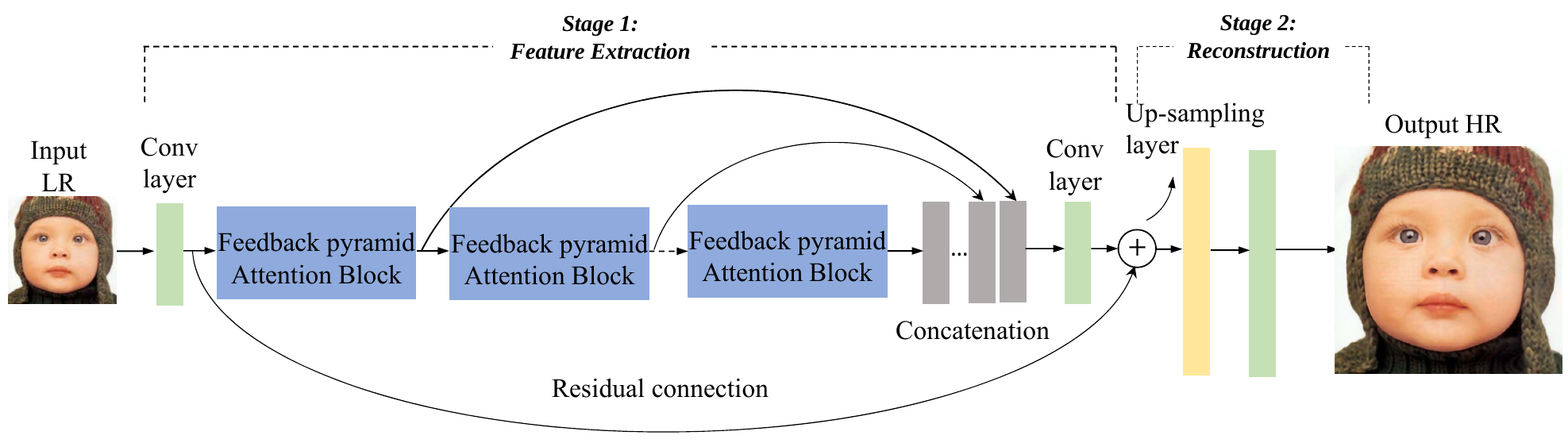}\\
  \caption{The architecture of our feedback pyramid attention network (FPAN). Our network consists of several feedback pyramid attention blocks. }\label{fig:architecture}
\end{figure*}

Although these feedforward methods achieve good results, the feedback mechanism \cite{felleman1991distributed, gilbert2007brain, hupe1998cortical, lamme2000the} has not been fully exploited in deep learning based super-resolution methods. In cognition theory, feedback connections that link the cortical visual areas can transfer response signals from high-level cortical areas to the current low-level neuron \cite{gilbert2007brain, hupe1998cortical}. Inspired by this mechanism, the feedback connections have been used in many deep neural networks \cite{carreira2016human, shrivastava2016contextual, zamir2017feedback}. The feedback connections can bring the semantic information from the higher layers back to refine the information of the previous layers in a top-down manner. Recently, Haris \emph{et al}. \cite{haris2018deep} introduced the feedback mechanism into an SR module to update feature errors iteratively. Liu \emph{et al}. \cite{liu2019hierarchical} proposed the new feedback version Hierarchical Back Projection Network (HBPN), applying the new up-sampling and down-sampling back projection block to return the LR and HR residual errors. To further improve the performance, Liu \emph{et al}. \cite{liu2019image} introduced the attention mechanism to the feedback block to refine the back projection network. Latterly, Li \emph{et al}. \cite{li2019feedback} employed the recurrent structure with feedback connections to correct the low-level representation information. CliqueNet constructed by Yang \emph{et al}. \cite{yang2018convolutional} that employed both forward and feedback dense connections in the same block, increases the discrimination of features. Later, Zhong \emph{et al}. \cite{zhong2018joint} introduced clique structures into super-resolution and obtained a better result. However, these extreme complex dense connections can increase the training difficulty of the network. Therefore, in this paper, we propose a novel simple feedback connection structure to promote the flow of information through the network.

\textbf{Attention mechanism.} In recent years, attention mechanisms have been successfully applied to a large number of tasks via rescaling the feature response to concentrate on the important components of the inputs \cite{hu2018squeeze, zhang2018image, dai2019second-order, wu2020multi-grained}. Hu \emph{et al}. \cite{hu2018squeeze} introduced the channel-wise attention mechanism, which proposed a squeeze-and-excitation block (SENet) to rescale channel-wise feature responses for image classification and achieved excellent results. Later, Zhang \emph{et al}. \cite{zhang2018image} directly integrated the squeeze-and-excitation block into the residual architecture to construct a residual channel attention network (RCAN) and achieved significant performance gains for image SR. Inspired by classic non-local mean, Wang \emph{et al}. \cite{wang2018non-local} introduced a non-local neural network that firstly incorporated non-local operations in deep convolutional neural networks for image recognition and video classification. In the non-local block, the spatial attention masks are obtained by computing the similarity matrix between each spatial position in the input feature map, then they adaptively modulate feature representations at a position by a weighted sum of the features at all positions based on the obtained attention masks. Considering that the non-local network (NL) calculates the correlation on the entire feature map, which will increase the computational complexity dramatically when the spatial dimensionality is large. In order to resolve this problem, Liu \emph{et al}. \cite{liu2018non-local} introduced a non-local recurrent network (NLRN) with limited neighborhood for image restoration. Dai \emph{et al}. \cite{dai2019second-order} adopted region-level non-local enhancement operations in second-order attention network (SAN). However, SAN \cite{dai2019second-order} and NLRN \cite{liu2018non-local} still suffer from a huge computational burden, which greatly hinders the use of non-local blocks. GCNet proposed by Cao \emph{et al}. \cite{cao2019gcnet:} surprisingly found that the attention maps of different query positions are almost the same. By simplifying the non-local block, GCNet has the advantage of the non-local block with capturing the long-range dependencies and the SENet with lightweight property.

\begin{figure*}
  \centering
  \includegraphics[height= 6in, width=\linewidth]{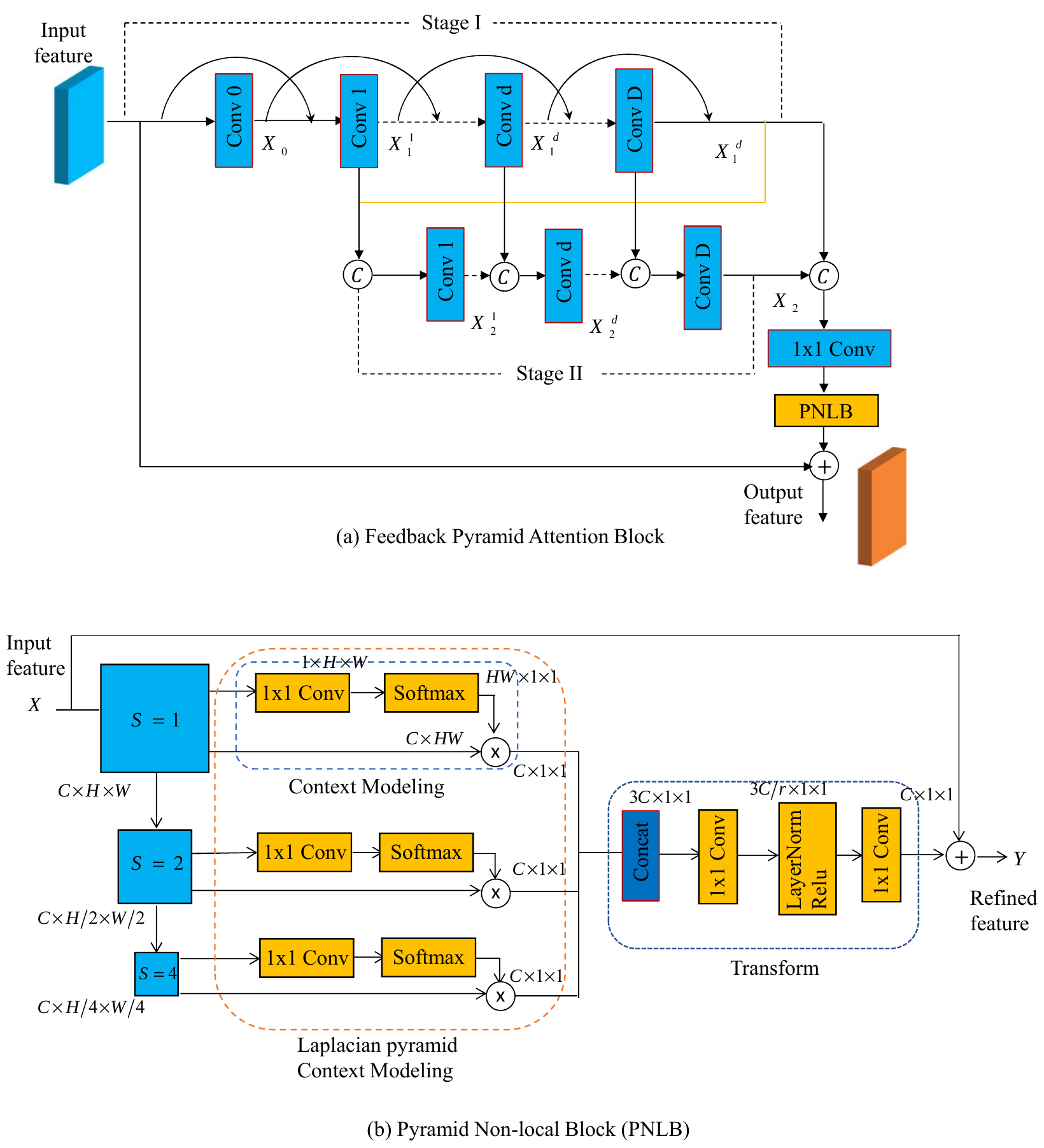}\\
  \caption{(a) The illustration of our feedback pyramid attention block (FPAB). (b) The details of the pyramid non-local block. $\bigotimes$ denotes matrix multiplication. $\oplus$ denotes broadcast element-wise addition.}\label{fig:FPAB}
\end{figure*}

In this paper, we propose a novel feedback pyramid attention network (termed FPAN) for SISR, which is shown in Fig. \ref{fig:architecture}. In the FPAN, we can not only enhance the representative ability of the proposed network but also improve the flow of information across the network by using a feedback connection structure. As a building block, feedback pyramid attention block (FPAB) is constructed by multiple forward skip connection layers and feedback skip connection layers, which can generate powerful high-level representations (illustrated in Fig. \ref{fig:FPAB}). The principle of our feedback scheme is that the output of each layer in the first stage is used as the input of the corresponding layer in the second stage to refine low-level representations. In each block, we further put forward to Laplacian pyramid attention to study non-local correlations at different scales, which is beneficial to enhance the discriminatory representation of the network (shown in Fig. \ref{fig:FPAB} (b)). The corresponding experiments show that our proposed method achieved favorable results in comparison with the state-of-the-art SR methods.

In summary, the main contributions of our method are as follows.




(1) We introduce a feedback pyramid attention network (FPAN) for SISR. Our FPAN can adaptively refine feature representations and capture richer context information for image SR.

(2) We propose a novel feedback connection mechanism, which not only efficiently improves the information flow, but also enhances high-level feature representation via multiple skip connections. Besides, the proposed pyramid non-local block can capture long-distance spatial contextual information at multiple scales, which efficiently and adaptively recalibrates feature presentations. In comparison with state-of-the-art methods, our method shows superiority in terms of model size and performance.

The rest of this paper is organized as follows. In Section \ref{sec:Methodology}, we analyze the proposed FPAN in detail. Experimental results with analysis are presented in Section \ref{sec:Experiments} to demonstrate the efficiency of our method. Finally, conclusions are provided in Section \ref{sec:Conclusion}.

\section{Proposed Method} \label{sec:Methodology}

In this section, we will introduce the proposed feedback pyramid attention networks in detail.

\subsection{Network Architecture}

The overall framework of our FPAN is illustrated in Fig. \ref{fig:architecture}, which is composed of two parts: 1) feature extraction part, and 2) reconstruction part. In this paper, $I^{LR}$ and  $I^{SR}$ represent the input LR image and output HR image of our FPAN. We firstly utilize one convolution layer to extract the initial features ${F_0}$ from the input LR.
\begin{equation}
    {F_{\rm{0}}} = {H_0}\left( {{I^{LR}}} \right),
\end{equation}
where ${H_0}$  denotes a convolution operation. ${F_0}$  is then fed into a series of feedback pyramid attention blocks (FPABs) to refine feature maps. Supposing G FPABs are used for further feature extraction, that is
\begin{equation}
    {F_G} = {M_G}\left( {{M_{G - 1}}\left( { \cdots {M_1}\left( {{F_0}} \right) \cdots } \right)} \right),
\end{equation}
where ${F_0}$ is the output of the first convolution layer ${H_0}$, ${M_g}\;(g = 1,2, \cdots ,G)$ denotes the $d$-th FPAB operation.

After acquiring a set of feature maps $\left[ {{F_G},{F_{G - 1}}, \cdots ,{F_1}} \right]$, we apply global feature fusion and global skip connection:
\begin{equation}
    {F_{GF}} = {F_{\rm{0}}} + {H_{hff}}\left[ {{F_G},{F_{G - 1}}, \cdots ,{F_1}} \right],
\end{equation}
where ${H_{hff}}$ is composed of hierarchical feature fusion ($1 \times 1$ convolution filter) and last feature extraction function (${\rm{3}} \times {\rm{3}}$  convolution filter), ${F_{GF}}$ is the output of the feature extraction part. Subsequently, we utilize a sub-pixel convolution layer \cite{shi2016real} followed by a convolution layer for reconstructing HR image.
\begin{equation}
    {I^{SR}} = {H_{recon}}\left( {{F_{GF}}} \right) = {H_{FPAN}}\left( {{I^{LR}}} \right),
\end{equation}
where ${H_{recon}}$ is the upscaling and reconstruction operation, ${H_{FPAN}}$ represents the function of our FPAN.

The FPAN is optimized with the ${L_{\rm{1}}}$ loss function \cite{zhang2018image, dai2019second-order, wu2020multi-grained} by minimizing the difference between the reconstructed image ${I^{SR}}$ and the ground-truth ${I^{HR}}$. Given a training dataset $\left\{ {I_i^{LR},I_i^{HR}} \right\}_{i = 1}^N$, where $N$ represents the number of the LR inputs and the corresponding HR inputs. The loss function is formulated as
\begin{equation}
    L(\theta ) = \frac{1}{N}\sum\limits_{i = 1}^N {{{\left| {{H_{FPAN}}\left( {I_i^{LR}} \right) - I_i^{HR}} \right|}_{\rm{1}}}},
\end{equation}
where $\theta$ denotes the parameter set of the FPAN. In the next subsection, we will introduce the proposed feedback pyramid attention block in detail.

\subsection{Feedback Pyramid Attention Block (FPAB)}

The structure of feedback pyramid attention block (FPAB) is shown in Fig. \ref{fig:FPAB}, the module mainly contains three parts: feedback connection structure, pyramid non-local block and local residual learning. Let ${F_{g - 1}}$ and ${F_{g}}$ denote the input and output of the $g$-th FPAB and ${F_{g}}$ can be obtained by
\begin{equation}
    {F_g} = {F_{g - 1}} + {f_{PN}}\left( {{f_{FC}}\left( {{F_{{\rm{g - 1}}}}} \right)} \right),
\end{equation}
where ${f_{FC}}$ and ${f_{PN}}$ are the proposed feedback connection structure and pyramid non-local block.

\subsubsection{Feedback Connection Structure}

Feedback mechanism is known as the efficient iterative procedure to refine low-level representations using high-level semantic information. Unlike the clique structure \cite{yang2018convolutional} that each layer is both the input and output of another one in the same block, we propose a novel feedback connection mechanism, where the output of each layer except the input layer in the first stage is used as the input of the corresponding layer in the second stage in the same block. As illustrated in Fig. \ref{fig:FPAB} (a), the proposed feedback structure contains two stages. At the beginning, the input layer ${F_{g - 1}}$ initializes all layers by simple feedforward skip connections. The outputs of two adjacent convolutional layers are fed to the subsequent layer in the first stage. The second stage begins to update the feature maps by building a skip connection between the corresponding layers of the two stages. Specifically, the proposed feedback connection mechanism can be presented as follows.

\noindent The initial stage is
\begin{equation}
    X_0^{}{\rm{ = }}\sigma \left( {{{\rm{W}}_0} * {F_{g - 1}}} \right).
\end{equation}

\noindent The first stage is
\begin{equation}
    X_1^i = \sigma \left( {W_1^i * [X_1^{i - 1},X_1^{i - 2}]} \right),
\end{equation}

\noindent and the second stage is
\begin{equation}
    X_2^i = \sigma \left( {W_2^i * [X_2^{i - 1},X_1^i]} \right),
\end{equation}
where $X_j^i$ denotes the feature maps of the $i$-th ($i \ge 1$) convolutional layer in the $j$-th ($j = 1,2$) stage. $X_1^{ - 1}{\rm{ = }}{{\rm{F}}_{g - 1}}$, $X_1^0 = {X_0}$. $X_2^0{\rm{ = }}{{\rm{X}}_1}$ is the output of the first stage. $ * $ is the convolutional operator. $\sigma $ represents the activation function (ReLU) \cite{nair2010rectified}. $\left[ { \cdot {\kern 1pt} ,{\kern 1pt}  \cdot } \right]$ refers to the concatenation operation. Then, the output of each stage is fed to the subsequent layers to enhance the information propagation in the network.

\subsubsection{Pyramid Non-local Block}

The aim of SR is to restore high frequency components of images. However, most SR methods recover image details blindly and lack abilities to identify high frequency regions. Thus, we consider if we can enhance the network sensitivity to high frequency information, thereby further improving the representational ability of the network, which is critical for SR. On the other hand, the convolution operation focuses only on local regions and cannot obtain long-dependencies at a time. In view of this, in this paper, we design a pyramid non-local block illustrated in Fig. \ref{fig:FPAB} (b), which can capture rich contextual information in the spatial dimensions and improve the discriminative ability of the model.

\paragraph{Non-local Module}

The non-local module \cite{wang2018non-local} can learn long-range relationships among any spatial positions and has been widely applied in visual tasks. In order to further improve the representation ability of the network, we will introduce a non-local module into the network to capture rich contextual information and enlarge the receptive field to the entire feature map. Let $X = \left\{ {{x_i}} \right\}_{i = 1}^N$ denotes the input of the non-local module, which has $C$ feature maps with size of $H \times W$. For the convenience of the calculation, we reshape X to $X\in \mathbb{R}^{C\times N}$, i.e., $N = H \times W$ and the spatial attention operation is defined as
\begin{equation}
    S = f\left( {\theta {{\left( X \right)}^T},\varphi \left( X \right)} \right),
\end{equation}
where $S\in \mathbb{R}^{N\times N}$, $\theta \left(  \cdot  \right)$ and $\varphi \left(  \cdot  \right)$ are two linear feature embeddings (e.g. $1 \times 1$ convolution operations). $f\left( { \cdot , \cdot } \right)$ is the similarity function that measures the correlations between any two positions in the embedding feature map. Specifically, most papers adopt softmax as a measure function:
\begin{equation}
    {s_{ij}} = f\left( {\theta {{\left( {{x_i}} \right)}^T},\varphi \left( {{x_j}} \right)} \right) = \frac{{\exp \left( {\theta \left( {{x_i}} \right) * \varphi \left( {{x_j}} \right)} \right)}}{{\sum\nolimits_{j = 1}^N {\exp \left( {\theta \left( {{x_i}} \right) * \varphi \left( {{x_j}} \right)} \right)} }},
\end{equation}
where ${s_{ij}}$ represents the similarity between the ${j^{th}}$ location ${x_j}$ and the ${i^{th}}$ location ${x_i}$ in the spatial dimension. Similarly, we also consider a linear embedding operation for input X: $g\left( X \right) = {W_g}X$ with the learned parameters ${W_g}$, which can be implemented as $1 \times 1$ convolution and we reshape it to $\mathbb{R}^{C\times N}$. Then, the final output ${y_i}$ at location $i$ as following:
\begin{equation}
    {y_i} = {W_y} * \sum\nolimits_{j = 1}^N {\left( {{s_{ij}} * g\left( {{x_j}} \right)} \right) + {x_i}},
\end{equation}
where $j$ is the index that enumerates all possible positions, capturing long-range contextual correlations across the whole feature map. ${W_y}$ is the learned weight parameter (initialized as zero) and the residual connection structure ``$ + {x_i}$''  guarantee the flow of the original information.

However, this non-local operation requires a lot of computational costs, especially when the spatial dimensionality is large. A global context block (GC) was presented in \cite{cao2019gcnet:}. The GC block simplifies the computational complexity of non-local module with SE block \cite{hu2018squeeze}, while enhancing the discriminative presentation of the network based on an independent attention map. The GC block mainly contains three steps, 1) adopt global attention pooling to implement context modeling, 2) transform operation to achieve channel-wise dependencies, 3) feature fusion to aggregate the global context feature. The global context block is defined as
\begin{equation}
    {y_i} = {x_i} + {W_{v2}}{\mathop{\rm Re}\nolimits} LU\left( {LN\left( {{W_{v1}}\sum\limits_{j = 1}^N {\frac{{{e^{{W_k}{x_j}}}}}{{\sum\limits_{m = 1}^N {{e^{{W_k}{x_m}}}} }}{x_j}} } \right)} \right),
\end{equation}
where ${W_k}$, ${W_{v1}}$, and ${W_{v2}}$ denote three linear feature embeddings (e.g. $1 \times 1$ convolution operations), ${\alpha _j} = \frac{{{e^{{W_k}{x_j}}}}}{{\sum\limits_{m = 1}^N {{e^{{W_k}{x_m}}}} }}$ represents the weight for context modeling, $LN\left(  \cdot  \right)$ is layer normalization \cite{ba2016layer}, $\delta \left(  \cdot  \right) = {W_{v2}}{\mathop{\rm Re}\nolimits} LU\left( {LN\left( {{W_{v1}}\left(  \cdot  \right)} \right)} \right)$ is the feature transform, and ``$ + $'' denotes the feature fusion.

\paragraph{Pyramid Non-local Mechanism}

To make full use of non-local power, pyramid attention is introduced to boost and exploit the relationship between the features at different scales. The corresponding architecture is illustrated in Fig. \ref{fig:FPAB} (b), formulated as
\begin{equation}
    \left\{ {\begin{array}{*{20}{l}}
{Y = X + PNLB\left( X \right)}\\
{PNLB\left( X \right) = \delta \left( {Concat\left( {P{A_1},P{A_2},P{A_4}} \right)} \right)}\\
{P{A_i} = \sum\limits_{j = 1}^N {a_j^i{x_j}} }\\
{{a^i} = soft\max \left( {W_k^i{X^i}} \right)}
\end{array}} \right.,
\end{equation}
where $PNLB\left(  \cdot  \right)$ denotes pyramid non-local operation, $i$ is the scale parameter of the Laplacian pyramid, $P{A_i}$ represents the context modeling module at the $i$-th scale. $Concat\left(  \cdot  \right)$ is a feature concatenation operation. The lightweight of the PNLB allows it to be plugged into multiple layers to better capture the long-range dependency with only a slight computational burden.

\section{Experimental Results and Analysis} \label{sec:Experiments}

\subsection{Datasets and Evaluation Metrics}

In the experiment, we adopt 800 high-quality images from DIV2K dataset \cite{timofte2017ntire} as the training set. Meanwhile, we use five standard benchmark datasets for test, including Set5 \cite{bevilacqua2012low}, Set14 \cite{zeyde2010single}, Bsd100 \cite{martin2001a}, Urban100 \cite{huang2015single} and Manga109 \cite{matsui2017sketch} with 5, 14, 100, 100 and 109 images for 3 upscaling factors, respectively. We carry out three experiments with three different degradation models, including bicubic (BI), blur-downscale (BD) and downscale-noise (DN) degradation models. The parameter settings of all degradation models are consistent with RDN \cite{zhang2018residual}. The SR images are first converted to YCbCr color space and evaluated by PSNR and SSIM \cite{wang2004image} on Y channel (i.e., luminance).

\begin{table*}
\newcommand{\tabincell}[2]{\begin{tabular}{@{}#1@{}}#2\end{tabular}}
\centering
\caption{Ablation study of different combinations in our method. The columns $P_0$ - $P_4$ correspond to different experiment configurations. We report the average PSNR on Set5 \cite{bevilacqua2012low} ( $\times {\rm{4}}$ ).}\label{ablation}
\begin{tabular}{l|C{1.2cm}|C{1.2cm}|C{1.2cm}|C{1.2cm}|C{1.2cm}}
\toprule
             & ${P_{\rm{0}}}$ & ${P_{\rm{1}}}$ & ${P_{\rm{2}}}$ & ${P_{\rm{3}}}$ & ${P_{\rm{4}}}$ \\
    \midrule
    Feedforward dense skip connections & & \checkmark & \checkmark  & \checkmark & \checkmark \\
    \hline
    Feedback skip connections & & & \checkmark  & \checkmark & \checkmark \\
    \hline
    Global context block ($S=$ 1) \cite{cao2019gcnet:} & & & & \checkmark \\
    \hline
    pyramid non-local block (PNLB, $S=$ 1, 2, 4) & & & & & \checkmark \\
    \hline
    PSNR (dB) & 32.33 & 32.35 & 32.43 & 32.49 & 32.55  \\
\bottomrule
 \end{tabular}
\end{table*}

\subsection{Implementation Details}

Data augmentation is performed on above 800 training images, which includes random rotating ${90^ \circ }$, ${\rm{18}}{{\rm{0}}^ \circ}$, ${\rm{27}}{{\rm{0}}^ \circ }$ and horizontal flipping. In each min-batch, 16 LR color patches with the size of $48 \times 48$ are randomly extracted as inputs. Our model is optimized with Adam \cite{kingma2014adam}, where the parameters are set to ${\beta _{\rm{1}}}{\rm{ = 0}}{\rm{.9}}$, ${\beta _{\rm{2}}}{\rm{ = 0}}{\rm{.999}}$, and $\varepsilon {\rm{ = 1}}{{\rm{0}}^{ - 8}}$. The learning rate is initialized to ${10^{ - 4}}$ and reduced to half every 200 epochs. We implement our model by utilizing PyTorch \cite{paszke2017automatic} with an Nvidia 2080Ti GPU.

\begin{figure*}
  \centering
  \subfigure[The relationship between parameters, PSNR, and S]{
  \includegraphics[width=0.45\linewidth]{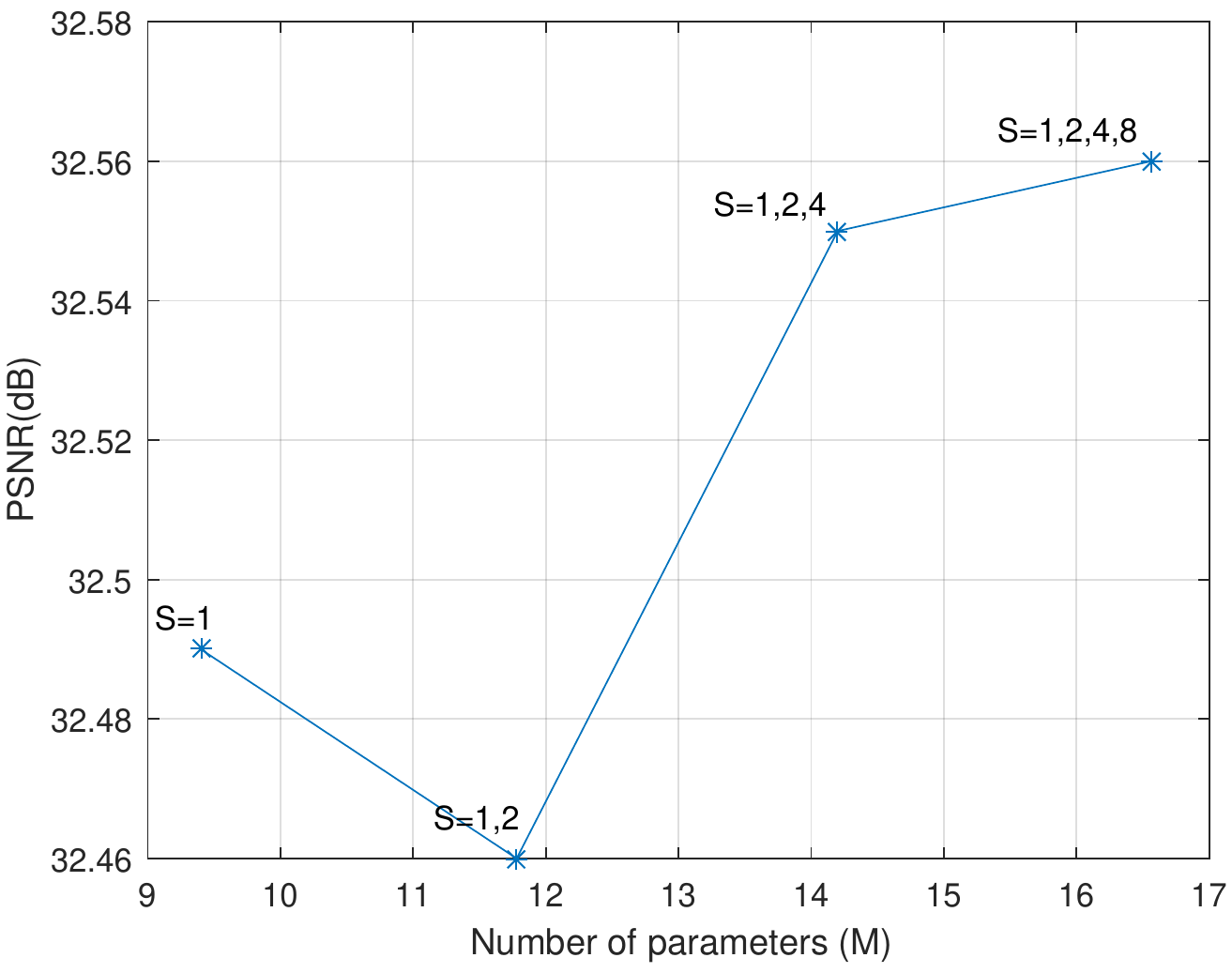}
  }
  \subfigure[The relationship between FLOPs, PSNR, and S]{
  \includegraphics[width=0.45\linewidth]{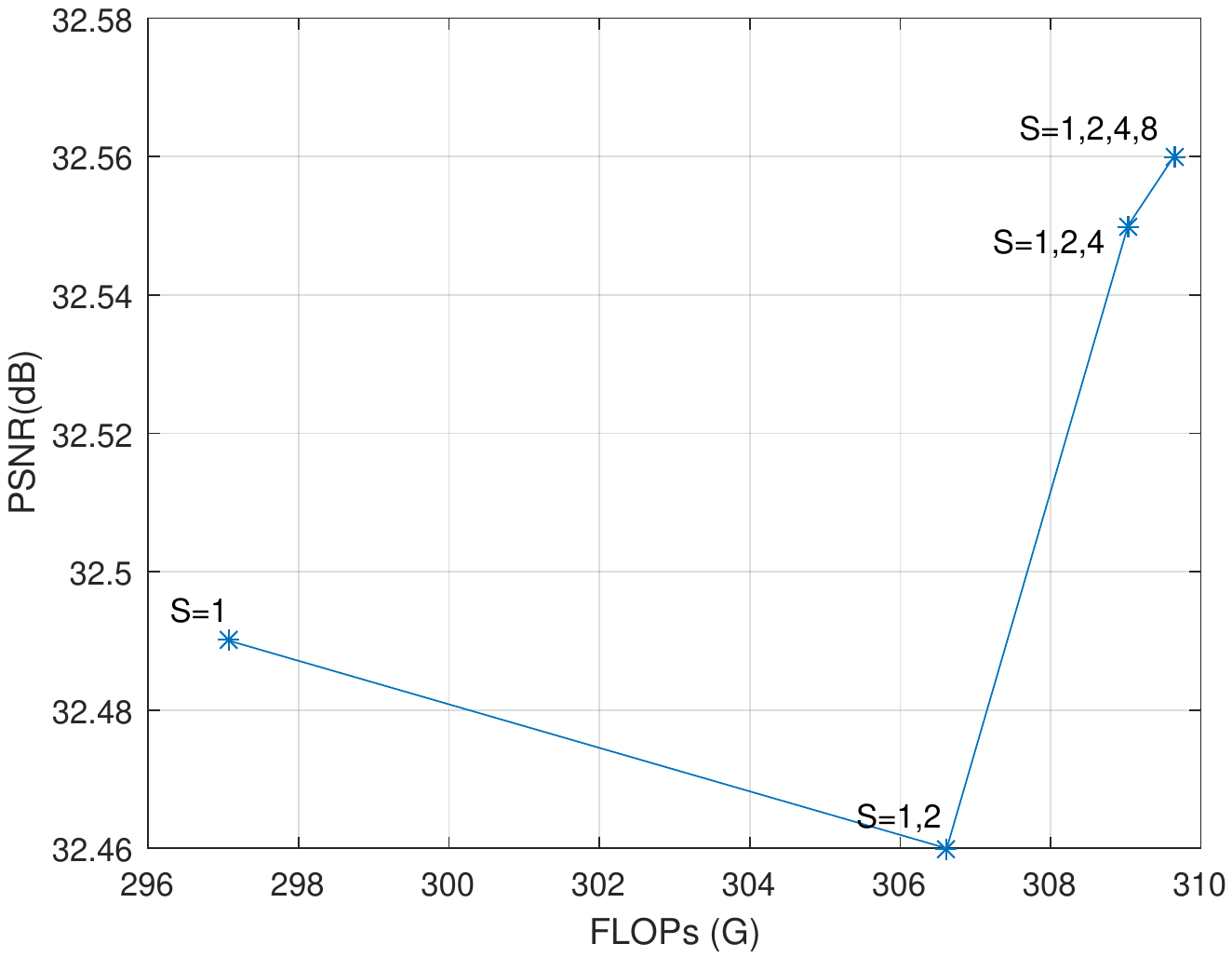}
  }
  \caption{Comparisons of the number of parameters, FLOPs and PSNR under different scale (S) settings in our method. }\label{fig:paraflops}
\end{figure*}

In the proposed network, we set all convolution layers to have 64 filters and the kernel size is ${\rm{3}} \times {\rm{3}}$ except for those   ${\rm{1}} \times {\rm{1}}$ convolutional layers in the pyramid non-local unit and feature fusion part. We adopt zero-padding strategy to keep the feature size fixed. In PNLB, we use ${\rm{6}} \times {\rm{6}}$ convolutional layer with two striding and two padding to perform down-sampling operations. In addition, the reduction ratio of the ${\rm{1}} \times {\rm{1}}$ convolutional layer is set to 16 in the pyramid non-local unit. In each feedback connection block, the number of convolutional layers is 8. For upscaling module, we follow \cite{lim2017enhanced, zhang2018image} and use ESPCNN \cite{shi2016real} to perform the upsampling operation. Meanwhile, the final convolution layer with 3 filters is used to produce the desired HR image.

\subsection{Ablation Analysis}

In this section, we investigate the effects of different components, including feedforward dense skip connections, feedback skip connections and pyramid attention. For a fair comparison, the baseline and its variants share the same experimental configuration and are trained from scratch.

 To demonstrate the effectiveness of the proposed feedback connection mechanism, we first compare the feedback network (${P_{\rm{2}}}$) with its feedforward counterpart (${P_{\rm{0}}}$, ${P_{\rm{1}}}$). As shown in Table \ref{ablation}, the baseline ${P_{\rm{0}}}$ (without the feedforward dense skip connections, the feedback skip connections and attention mechanism) performance is relatively low, its PSNR only reaches 32.33 dB on Set5 $\left( { \times 4} \right)$. Then, ${P_{\rm{1}}}$ that adds the feedforward dense skip connections obtain 32.35 dB. After integrating the feedback skip connections, the performance can be improved from 32.35 dB to 32.43 dB. Meanwhile, as discussed above, we further show the effect of attention. It can be seen that the network with attention (${P_{\rm{3}}}$ and ${P_{\rm{4}}}$) perform better than those without attention (from ${P_{\rm{0}}}$ to ${P_{\rm{2}}}$). These observations indicate that the attention mechanism can adaptively exploit more important information for SR. Specifically, when we compare the global context block \cite{cao2019gcnet:} (${P_{\rm{3}}}$, $S=$ 1) and pyramid non-local block (${P_{\rm{4}}}$, $S=$ 1, 2, 4), we can see that our pyramid attention mechanism can achieve better performance than the network with the global context block (i.e., 32.55 dB v.s. 32.49 dB). In addition, we also show some other quantitative comparison results, including: the number of parameters and the number of floating-point operations (FLOPs), which are illustrated in Fig.~\ref{fig:paraflops}. FLOPs is computed with $512 \times 512$ HR image at $ \times {\rm{4}}$ scaling factor. It can be find that although more pyramid scales can improve PSNR, the number of parameters and PLOPs will also increase. To achieve a better tradeoff, we set the pyramid scale to 1, 2, 4. These comparisons consistently demonstrate the superiority of our proposed FPAN.

\subsection{Results with Bicubic Degradation (BI)}

To verify the effectiveness of our FPAN, we compare our FPAN with some other state-of-the-art SR methods: Bicubic, SRCNN \cite{dong2015image}, VDSR \cite{kim2016accurate}, DRCN \cite{kim2016deeply}, LapSRN \cite{lai2017deep}, EDSR \cite{lim2017enhanced}, DBPN \cite{haris2018deep} and SRFBN \cite{li2019feedback}. As in \cite{lim2017enhanced, zhang2018residual, zhang2018image}, we also adopt self-ensemble strategy to further improve our FPAN and denote the self-ensembled FPAN as FPAN+. Table \ref{comp_with_others_BI} shows all the quantitative results for various scale factors. In comparison with these methods, our FPAN+ achieves the best performance on all the datasets with various scale factors. Without self-ensemble, our FPAN also obtains comparable performance and outperforms almost all the other methods. Although there is no absolute advantage compared to some state-of-the-art methods, our FPAN has fewer layers and parameters than EDSR \cite{lim2017enhanced} (11.7 M v.s. 43 M) and employs fewer training images than DBPN \cite{haris2018deep} and SRFBN \cite{li2019feedback} (DIV2K v.s. DIV2K + Flickr2K + ImageNet v.s. DIV2K + Flickr2K).

\begin{figure*}
  \centering
  \includegraphics[width=\linewidth]{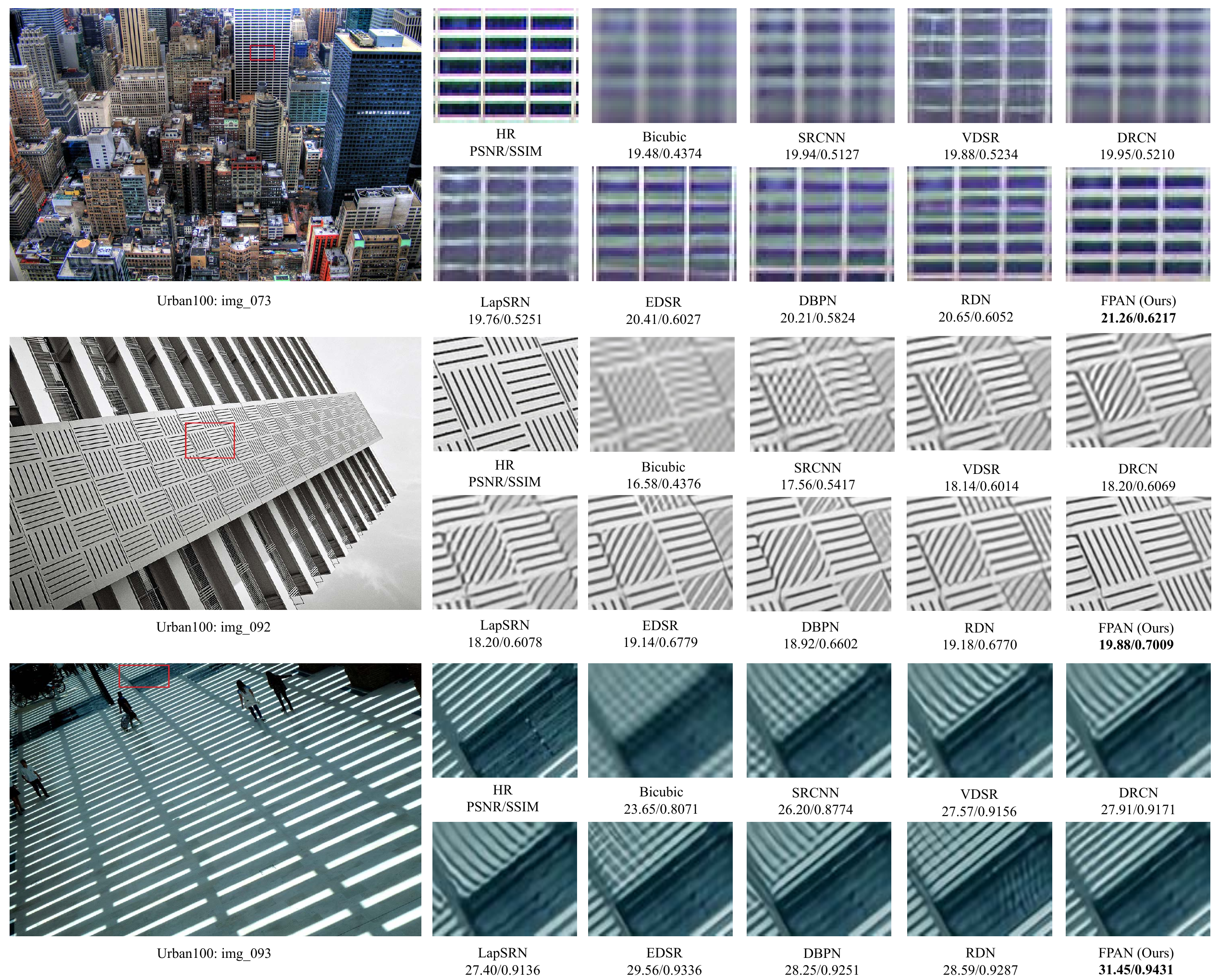}\\
  \caption{Visual quality for $ \times {\rm{4}}$ SR with BI model on Urban100 dataset \cite{huang2015single}. The best results are highlighted.
}\label{fig:img073092093-Urban100}
\end{figure*}

To further illustrate the superiority of our FPAN, in Fig.~\ref{fig:img073092093-Urban100}, we show visual results with an upscaling factor $\times {\rm{4}}$ for these images from Urban100 dataset. From that, we can see that our method yields the best visual results among all existing compared methods. Using ``$image\_092$'' as an example, we can see that the Bicubic interpolation generates heavy blurring artifacts along the edges and visually displeasing textures. Later, some developed methods can recover some structure, but still suffer from the wrong texture direction and fail to recover more accurate image details. However, the proposed FPAN alleviates distorted lines and produces more faithful textures and structures. The reason is that FPAN can use high-level information to refine the low-level feature information to achieve self-correction. In addition, pyramid attention mechanism is used to further enhance the discriminative representation of the network, thus generating a better SR result.

\begin{figure*}
  \centering
  \includegraphics[width=\linewidth]{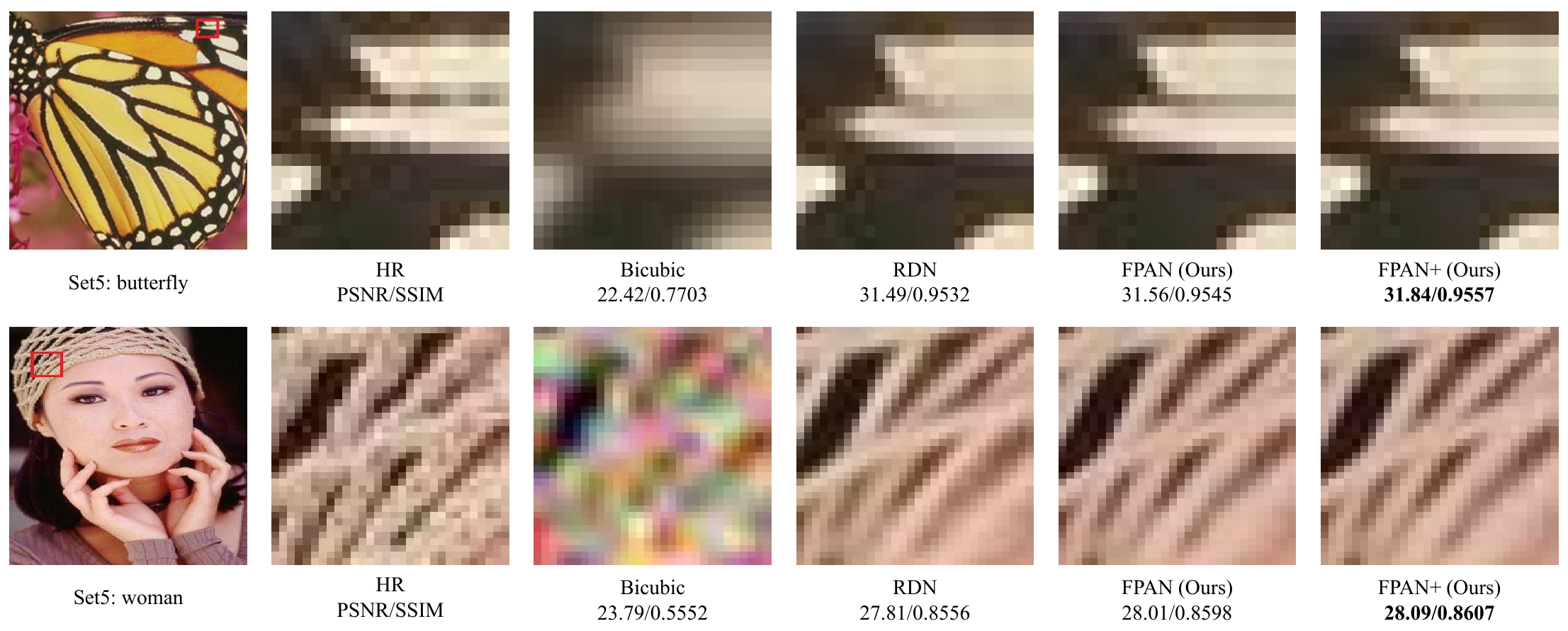}\\
  \caption{
  Visual quality for $ \times {\rm{3}}$ SR with BD and DN models on Set5 dataset \cite{bevilacqua2012low}. The best results are highlighted.
 }\label{fig:img-Set5}
\end{figure*}

\subsection{Results with BD and DN Degradations}

Following \cite{zhang2018residual, li2019feedback}, we also show SR results with BD and DN degradation models. We compare our FPAN with some state-of-the-art methods: SRCNN \cite{dong2015image}, FSRCNN \cite{dong2016accelerating}, VDSR \cite{kim2016accurate}, ”IRCNN\_G“ \cite{zhang2017learning}, ”IRCNN\_C“ \cite{zhang2017learning}, RDN \cite{zhang2018residual}, SRFBN \cite{li2019feedback}. As shown in Table \ref{comp_with_others_BD}, our FPAN and FPAN+ generate better performance than other methods on almost all quantitative results with scaling factor $ \times {\rm{3}}$. In particular, for the BD degradation model, our method achieves better performance than other state-of-the-art methods even without self-ensemble.

In Fig. \ref{fig:img-Set5}, we show visual comparisons with BD and DN degradation models. The recovered results from other methods are blurry. In contrast, our FPAN and FPAN+ can alleviate the distortion and produce more faithful details in SR results. It also verifies that our FPAN has the ability to capture the important discriminative features for BD and DN degradation models.

\subsection{Model Size Comparison}

\begin{figure}
  \centering
  \includegraphics[width=\linewidth]{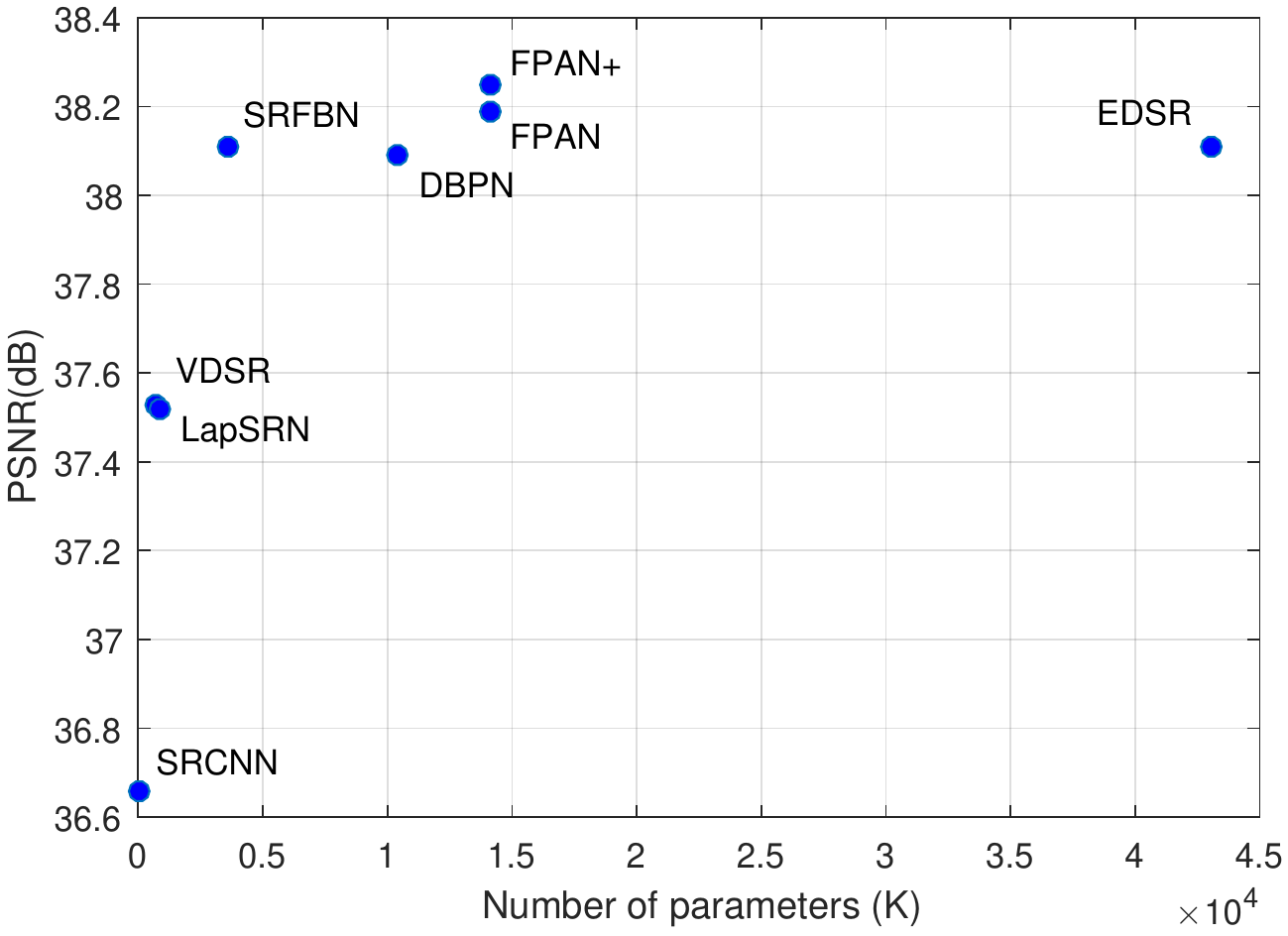}\\
  \caption{PSNR vs. the number of parameters. Results are evaluated on Set 5 \cite{bevilacqua2012low} for up-scaling factor $ \times {\rm{2}}$.}\label{fig:san}
\end{figure}

In Fig. \ref{fig:san}, we illustrate the comparisons about model size and performance on Set 5 with scale factor $ \times {\rm{2}}$. Our FPAN has much fewer parameters than EDSR, but obtains higher performance. In comparison with DBPN and SRFBN, our FPAN and FPAN+ obtain better results with a relatively large model. It implies that our FPAN has a good tradeoff between model size and performance.

\begin{table*}
\newcommand{\tabincell}[2]{\begin{tabular}{@{}#1@{}}#2\end{tabular}}
\centering
\caption{Quantitative results with BI degradation module. Average PSNR/SSIM comparison results for scaling factor $ \times {\rm{2}}$, $ \times {\rm{3}}$ and $ \times {\rm{4}}$.}\label{comp_with_others_BI}
\begin{tabular}{L{1.8cm}|C{0.8cm}|C{0.8cm}C{0.8cm}|C{0.8cm}C{0.8cm}|C{0.8cm}C{0.8cm}|C{0.8cm}C{0.8cm}|C{0.8cm}C{0.8cm}}
\toprule
\multicolumn{1}{c}{} & \multicolumn{1}{c}{} &
\multicolumn{2}{c}{Set5} & \multicolumn{2}{c}{Set14} & \multicolumn{2}{c}{Bsd100} & \multicolumn{2}{c}{Urban100} & \multicolumn{2}{c}{Manga109}\\
    Methods & Scale & PSNR & SSIM & PSNR & SSIM & PSNR & SSIM & PSNR & SSIM & PSNR & SSIM \\
\midrule
    Bicubic & x2 & 33.66 & 0.9299 & 30.24 & 0.8688 & 29.56 & 0.8431 & 26.88 & 0.8403 & 30.80 & 0.9339 \\
    SRCNN \cite{dong2015image} & 2 & 36.66 & 0.9542 & 32.45 & 0.9067 & 31.36 & 0.8879 & 29.50 & 0.8946 & 35.60 & 0.9663 \\
    VDSR \cite{kim2016accurate} & 2 & 37.53 & 0.9590 & 33.05 & 0.9130 & 31.90 & 0.8960 & 30.77 & 0.9140 & 37.22 & 0.9750 \\
    DRCN \cite{kim2016deeply} & 2 & 37.63 & 0.9588 & 33.04 & 0.9118 & 31.85 & 0.8942 & 30.75 & 0.9133 & 37.63 & 0.9723 \\
    LapSRN \cite{lai2017deep} & 2 & 37.52 & 0.9591 & 33.08 & 0.9130 & 31.08 & 0.8950 & 30.41 & 0.9101 & 37.27 & 0.9740 \\
    EDSR \cite{lim2017enhanced} & 2 & 38.11 & 0.9602 & 33.92 & 0.9195 & 32.32 & 0.9013 & 32.93 & 0.9351 & 39.10 & 0.9773 \\
    DBPN \cite{haris2018deep} & 2 & 38.09 & 0.9600 & 33.85 & 0.9190 & 32.27 & 0.9000 & 32.55 & 0.9324 & 38.89 & 0.9775 \\
    SRFBN \cite{li2019feedback} & 2 & 38.11 & 0.9609 & 33.82 & 0.9196 & 32.29 & 0.9010 & 32.62 & 0.9328 & 39.08 & 0.9779 \\
    FPAN (ours) & 2 & 38.19 & 0.9612 & 33.88 & 0.9210 & 32.30 & 0.9012 & 32.72 & 0.9339 & 39.03 & 0.9772 \\
    FPAN+ (ours) & 2 & \textbf{38.25} & \textbf{0.9614} & \textbf{33.93} & \textbf{0.9210} & \textbf{32.35} & \textbf{0.9017} & \textbf{32.94} & \textbf{0.9357} & \textbf{39.28} & \textbf{0.9780} \\
\midrule
    Bicubic & 3 & 30.39 & 0.8682 & 27.55 & 0.7742 & 27.21 & 0.7385 & 24.46 & 0.7349 & 26.95 & 0.8556 \\
    SRCNN \cite{dong2015image} & 3 & 32.75 & 0.9090 & 29.30 & 0.8215 & 28.41 & 0.7863 & 26.24 & 0.7989 & 30.48 & 0.9117 \\
    VDSR \cite{kim2016accurate} & 3 & 33.67 & 0.9210 & 29.78 & 0.8320 & 28.83 & 0.7990 & 27.14 & 0.8290 & 32.01 & 0.9340 \\
    DRCN \cite{kim2016deeply} & 3 & 33.82 & 0.9226 & 29.76 & 0.8311 & 28.80 & 0.7963 & 27.15 & 0.8276 & 32.31 & 0.9328 \\
    LapSRN \cite{lai2017deep} & 3 & 33.82 & 0.9227 & 29.87 & 0.8320 & 28.82 & 0.7980 & 27.07 & 0.8280 & 32.21 & 0.9350 \\
    EDSR \cite{lim2017enhanced} & 3 & 34.65 & 0.9280 & 30.52 & 0.8462 & 29.25 & 0.8093 & 28.80 & 0.8653 & 34.17 & 0.9476 \\
    SRFBN \cite{li2019feedback} & 3 & 34.70 & 0.9292 & 30.51 & 0.8461 & 29.24 & 0.8084 & 28.73 & 0.8641 & 34.18 & 0.9481 \\
    FPAN (ours) & 3 & 34.62 & 0.9291 & 30.55 & 0.8467 & 29.24 & 0.8090 & 28.73 & 0.8642 & 34.14 & 0.9481 \\
    FPAN+ (ours) & 3 & \textbf{34.72} & \textbf{0.9298} & \textbf{30.66} & \textbf{0.8480} & \textbf{29.30} & \textbf{0.8102} & \textbf{28.98} & \textbf{0.8677} & \textbf{34.49} & \textbf{0.9497} \\
    \midrule
    Bicubic & 4 & 28.42 & 0.8104 & 26.00 & 0.7027 & 25.96 & 0.6675 & 23.14 & 0.6577 & 24.89 & 0.7866 \\
    SRCNN \cite{dong2015image} & 4 & 30.48 & 0.8628 & 27.50 & 0.7513 & 26.90 & 0.7101 & 24.52 & 0.7221 & 27.58 & 0.8555 \\
    VDSR \cite{kim2016accurate} & 4 & 31.35 & 0.8830 & 28.02 & 0.7680 & 27.29 & 0.7251 & 25.18 & 0.7540 & 28.83 & 0.8870 \\
    DRCN \cite{kim2016deeply} & 4 & 31.53 & 0.8854 & 28.02 & 0.7670 & 27.23 & 0.7233 & 25.14 & 0.7510 & 28.98 & 0.8816 \\
    LapSRN \cite{lai2017deep} & 4 & 31.54 & 0.8850 & 28.19 & 0.7720 & 27.32 & 0.7270 & 25.21 & 0.7560 & 29.09 & 0.8900 \\
    EDSR \cite{lim2017enhanced} & 4 & 32.46 & 0.8968 & 28.80 & 0.7876 & 27.71 & 0.7420 & 26.64 & 0.8033 & 31.02 & 0.9148 \\
    DBPN \cite{haris2018deep} & 4 & 32.47 & 0.8980 & 28.82 & 0.7860 & 27.72 & 0.7400 & 26.38 & 0.7946 & 30.91 & 0.9137 \\
    SRFBN \cite{li2019feedback} & 4 & 32.47 & 0.8983 & 28.81 & 0.7868 & 27.72 & 0.7409 & 26.60 & 0.8015 & 31.15 & 0.9160 \\
    FPAN (ours) & 4 & 32.48 & 0.8984 & 28.78 & 0.7867 & 27.71 & 0.7412 & 26.61 & 0.8025 & 30.99 & 0.9144 \\
    FPAN+ (ours) & 4 & \textbf{32.62} & \textbf{0.8999} & \textbf{28.90} & \textbf{0.7889} & \textbf{27.77} & \textbf{0.7427} & \textbf{26.82} & \textbf{0.8070} & \textbf{31.36} & \textbf{0.9178} \\
\bottomrule
 \end{tabular}
\end{table*}

\begin{table*}
\newcommand{\tabincell}[2]{\begin{tabular}{@{}#1@{}}#2\end{tabular}}
\centering
\caption{Quantitative results with BD and DN degradation modules. Average PSNR/SSIM comparison results for scaling factor $ \times 3$.}\label{comp_with_others_BD}
\begin{tabular}{L{1.8cm}|C{0.8cm}|C{0.8cm}C{0.8cm}|C{0.8cm}C{0.8cm}|C{0.8cm}C{0.8cm}|C{0.8cm}C{0.8cm}|C{0.8cm}C{0.8cm}}
\toprule
\multicolumn{1}{c}{} & \multicolumn{1}{c}{} &
\multicolumn{2}{c}{Set5} & \multicolumn{2}{c}{Set14} & \multicolumn{2}{c}{Bsd100} & \multicolumn{2}{c}{Urban100} & \multicolumn{2}{c}{Manga109}\\
    Methods & Model & PSNR & SSIM & PSNR & SSIM & PSNR & SSIM & PSNR & SSIM & PSNR & SSIM \\
\midrule
\midrule
    \multirow{2}{*}{Bicubic} & BD & 28.78 & 0.8308 & 26.38 & 0.7271 & 26.33 & 0.6918 & 23.52 & 0.6862 & 25.46 & 0.8149 \\
      & DN & 24.01 & 0.5369 & 22.87 & 0.4724 & 22.92 & 0.4449 & 21.63 & 0.4687 & 23.01 & 0.5381 \\
\midrule
    \multirow{2}{*}{SRCNN \cite{dong2015image}} & BD & 32.05 & 0.8944 & 28.80 & 0.8074 & 28.13 & 0.7736 & 25.70 & 0.7770 & 29.47 & 0.8924 \\
     & DN & 25.01 & 0.6950 & 23.78 & 0.5898 & 23.76 & 0.5538 & 21.90 & 0.5737 & 23.75 & 0.7148 \\
\midrule
    \multirow{2}{*}{FSRCNN \cite{dong2016accelerating}} & BD & 26.23 & 0.8124 & 24.44 & 0.7106 & 24.86 & 0.6832 & 22.04 & 0.6745 & 23.04 & 0.7927 \\
    & DN & 24.18 & 0.6932 & 23.02 & 0.5856 & 23.41 & 0.5556 & 21.15 & 0.5682 & 22.39 & 0.7111 \\
\midrule
    \multirow{2}{*}{VDSR \cite{kim2016accurate}} & BD & 33.25 & 0.9150 & 29.46 & 0.8244 & 28.57 & 0.7893 & 26.61 & 0.8136 & 31.06 & 0.9234 \\
    & DN & 25.20 & 0.7183 & 24.00 & 0.6112 & 24.00 & 0.5749 & 22.22 & 0.6096 & 24.20 & 0.7525 \\
\midrule
    \multirow{2}{*}{“IRCNN\_G“ \cite{zhang2017learning}} & BD & 33.38 & 0.9182 & 29.63 & 0.8281 & 28.65 & 0.7922 & 26.77 & 0.8154 & 31.15 & 0.9245 \\
    & DN & 25.70 & 0.7379 & 24.45 & 0.6305 & 24.28 & 0.5900 & 22.90 & 0.6429 & 24.88 & 0.7765 \\
\midrule
    \multirow{2}{*}{”IRCNN\_C“ \cite{zhang2017learning}} & BD & 33.17 & 0.9157 & 29.55 & 0.8271 & 28.49 & 0.7886 & 26.47 & 0.8081 & 31.13 & 0.9236 \\
    & DN & 27.48 & 0.7925 & 25.92 & 0.6932 & 25.55 & 0.6481 & 23.93 & 0.6950 & 26.07 & 0.8253 \\
\midrule
    \multirow{2}{*}{RDN \cite{zhang2018residual}} & BD & 34.58 & 0.9280 & 30.53 & 0.8447 & 29.23 & 0.8079 & 28.46 & 0.8582 & 33.97 & 0.9465 \\
    & DN & 28.47 & 0.8151 & 26.60 & 0.7101 & 25.93 & 0.6573 & 24.92 & 0.7364 & 28.00 & 0.8591 \\
\midrule
    \multirow{2}{*}{SRFBN \cite{li2019feedback}} & BD & 34.66 & 0.9283 & 30.48 & 0.8439 & 29.21 & 0.8069 & 28.48 & 0.8581 & 34.07 & 0.9466 \\
    & DN & 28.53 & 0.8182 & 26.60 & \textbf{0.7144} & 25.95 & \textbf{0.6625} & 24.99 & 0.7424 & 28.02 & 0.8618 \\
\midrule
    \multirow{2}{*}{FPAN (ours)} & BD & 34.62 & 0.9286 & 30.60 & 0.8455 & 29.27 & 0.8089 & 28.65 & 0.8616 & 34.36 & 0.9479 \\
    & DN & 28.53 & 0.8182 & 26.63 & 0.7113 & 25.95 & 0.6603 & 25.07 & 0.7427 & 28.07 & 0.8613 \\
\midrule
    \multirow{2}{*}{FPAN+ (ours)} & BD & \textbf{34.72} & \textbf{0.9292} & \textbf{30.71} & \textbf{0.8474} & \textbf{29.34} & \textbf{0.8100} & \textbf{28.88} & \textbf{0.8649} & \textbf{34.71} & \textbf{0.9496} \\
    & DN & \textbf{28.62} & \textbf{0.8193} & \textbf{26.69} & 0.7128 & \textbf{25.98} & 0.6612 & \textbf{25.19} & \textbf{0.7460} & \textbf{28.24} & \textbf{0.8638} \\
\bottomrule
 \end{tabular}
\end{table*}

\section{Conclusion} \label{sec:Conclusion}

In this paper, we propose a deep feedback pyramid attention networks (FPAN) for image SR. Specially, the proposed feedback connection structure allows FPAN to not only refine low-level representation with high-level information, but also effectively boost the flow of information as well as the feature reuse. In addition, we introduce a pyramid non-local block to capture the long-distance dependencies in multiple scales and make the network concentrate on learning important high-frequency information. Extensive evaluations on SR with BI, BD and DN models show the superior performance of our FPAN in terms of quantitative and visual results.

\appendices

\ifCLASSOPTIONcaptionsoff
  \newpage
\fi

\bibliographystyle{IEEEtran}   
\bibliography{egbib}   

\end{document}